\documentstyle[11pt]{article}
\input epsf                                
\catcode`\@=11
\@addtoreset{equation}{section}
\def\theequation{\arabic{section}.\arabic{equation}}
\def\appendix{\renewcommand{\thesection}{\Alph{section}}\setcounter{section}{0}
              \renewcommand{\theequation}
            {\mbox{\Alph{section}.\arabic{equation}}}\setcounter{equation}{0}}
\def\maketitle{\thispagestyle{empty}\setcounter{page}0\newpage
                \renewcommand{\thefootnote}{\arabic{footnote}}
                  \setcounter{footnote}0}
\renewcommand{\thanks}[1]{\renewcommand{\thefootnote}{\fnsymbol{footnote}}
               \footnote{#1}\renewcommand{\thefootnote}{\arabic{footnote}}}
\newcommand{\preprint}[1]{\hfill{\sl preprint - #1}\par\bigskip\par\rm}
\renewcommand{\title}[1]{\begin{center}\Large\bf #1\end{center}\rm\par\bigskip}
\renewcommand{\author}[1]{\begin{center}\Large #1\end{center}}
\newcommand{\address}[1]{\begin{center}\large #1\end{center}}

\def\dinfn{\smallskip $^2 $Dipartimento di Fisica, Universit\`a di Trento\\
                           and Istituto Nazionale di Fisica Nucleare,\\
                                   Gruppo Collegato di Trento, Italia}

\def\Idinfn{\address{\dinfn}}
\def\dinbcn{\smallskip $^1 $Consejo Superior de Investigaciones 
Cient\'{\i}ficas, \\ IEEC,
Edifici Nexus 104,
Gran Capit\`a 2-4, 08034 Barcelona, Spain\\
                           and
Departament ECM and IFAE, Facultat de F\'{\i}sica, \\
Universitat de Barcelona, Diagonal 647, 08028 Barcelona, Spain }
\def\Idinbcn{\address{\dinbcn}}
\newcommand{\zmail}[1]{e-mail: \sl #1@zeta.ecm.ub.es\rm}
\newcommand{\fzmail}[1]{\thanks{\zmail{#1}}}
\newcommand{\email}[1]{e-mail: \sl #1@science.unitn.it\rm}
\newcommand{\femail}[1]{\thanks{\email{#1}}}
\newcommand{\pacs}[1]{\smallskip\noindent{\sl PACS numbers:
                       \hspace{0.3cm}#1}\par\bigskip\rm}
\def\babs{\hrule\par\begin{description}\item{Abstract: }\it}
\def\eabs{\par\end{description}\hrule\par\medskip\rm}
\renewcommand{\date}[1]{\par\bigskip\par\sl\hfill #1\par\medskip\par\rm}
\newcommand{\ack}[1]{\par\section*{Acknowledgments} #1}
\newcommand{\s}[1]{\section{#1}}

\newcommand{\ca}[1]{{\cal #1}}         
\def\hs{\qquad}               
\def\nn{\nonumber}            
\def\beq{\begin{eqnarray}}    
\def\eeq{\end{eqnarray}}      
\def\ap{\left.}               
\def\at{\left(}               
\def\aq{\left[}               
\def\cp{\right.}              
\def\ct{\right)}              
\def\cq{\right]}              
\def\R{{\hbox{{\rm I}\kern-.2em\hbox{\rm R}}}}   
\def\H{{\hbox{{\rm I}\kern-.2em\hbox{\rm H}}}}   
\def\N{{\hbox{{\rm I}\kern-.2em\hbox{\rm N}}}}   
\def\C{{\ \hbox{{\rm I}\kern-.6em\hbox{\bf C}}}} 
\def\Z{{\hbox{{\rm Z}\kern-.4em\hbox{\rm Z}}}}   
\def\ii{\infty}                                  
\newcommand{\fr}[2]{\mbox{$\frac{#1}{#2}$}}      
\def\tr{\mathop{\rm tr}\nolimits}                  
\def\Tr{\mathop{\rm Tr}\nolimits}                  
\def\Res{\mathop{\rm Res}\nolimits}                
\renewcommand{\Re}{\mathop{\rm Re}\nolimits}       
\def\lap{\Delta}                                   
\def\al{\alpha}

\def\ga{\gamma}
\def\de{\delta}

\def\ze{\zeta}

\def\la{\lambda}

\def\Ga{\Gamma}

\begin{document}

\preprint{UTF 394}

\title{Zeta-function Regularization, the  Multiplicative Anomaly \\
and the Wodzicki Residue}

\author{Emilio Elizalde$^{1,}$\fzmail{eli},
Luciano Vanzo$^{2,}$\femail{vanzo} and
Sergio Zerbini$^{2,}$\femail{zerbini}
}
\Idinbcn
\Idinfn


\babs
The multiplicative anomaly associated with the zeta-function
regularized determinant is computed  for the Laplace-type operators
$L_1=-\lap+V_1$ and  $L_2=-\lap+V_2$, with $V_1$,  $V_2$ constant, in a
$D$-dimensional  compact smooth manifold $ M_D$,
making use of several results due to Wodzicki and
 by direct calculations in some explicit examples. It is found that
the multiplicative anomaly is vanishing  for $D$ odd and for $D=2$.
An application to the one-loop effective potential of the $O(2)$
self-interacting scalar model is outlined.

\eabs

\pacs{04.62.+v, 04.70.Dy}

\noindent Running title: Zetas, multiplicative anomaly and Wodzicki residue


\s{Introduction}
\label{Form}

Within the one-loop or external field approximation,
the importance of zeta-function regularization for functional 
determinants, as introduced in
\cite{ray71-7-145}, is well known, as a
powerful tool to deal with the ambiguities (ultraviolet divergences) present
in  relativistic quantum field theory (see for example
\cite{eliz94b}-\cite{byts96-266-1}).
It permits to give a meaning, in the sense of analytic continuation,
to the determinant of a differential operator which, as the product of its
eigenvalues, is formally divergent.
For the sake of simplicity we shall here restrict ourselves to scalar
fields. The one-loop Euclidean partition function, regularised by
zeta-function techniques, reads \cite{hawk77-55-133}
\beq
\ln Z=-\frac{1}{2}\ln\det \frac{L_D}{\mu^2}
=\frac{1}{2}\ze'(0|L_D)+\frac{1}{2}\ze(0|L_D)\ln\mu^2
\nn\:,\eeq
where $\ze(s|L_D)$ is the zeta function related to $L_D$
---typically an elliptic differential operator of
second order---
$\ze'(0|L_D)$ its derivative with respect to $s$, and
$\mu^2$ a renormalization scale.
The fact is used that the analytically continued zeta-function is
 generally regular at $s=0$,
and thus its derivative is well defined.

When the manifold is smooth and compact, the spectrum is discrete and one has
\beq
\ze(s|L_D)=\sum_i\la_i^{-2s}
\:,\nn\eeq
$\la^2_i$ being the eigenvalues of $L_D$. As a result,
one can make use of the relationship between the zeta-function and the
heat-kernel trace via the the Mellin transform and its inverse.
For $\Re s>D/2$, one can write
\beq
\ze(s|L_D)=\Tr L_D^{-s}
=\frac{1}{\Ga(s)}\int_0 ^\ii t^{s-1}
\:K(t|L_D)\:dt
\,,\label{mt}\eeq
\beq
K(t|L_D)=\frac{1}{2\pi i}
\int_{\Re s >D/2} t^{-s}\:\Ga(s)\ze(s|L_D)\:ds
\:,\label{minv}
\eeq
where $K(t|L_D)=\Tr\exp(-t L_D)$ is the heat operator.
The previous relations are valid also in the presence of
zero modes, with the replacement
$K(t|L_D)\longrightarrow K(t|L_D)-P_0$,
$P_0$ being the projector onto the zero modes.

A heat-kernel expansion argument leads to the meromorphic structure of
$\zeta(s|L_D)$ and,
as we have anticipated, it is found that the analytically continued zeta-function is regular
at $s=0$ and thus its derivative is well defined.
Furthermore, in practice all the operators may be considered to be
trace-class. In fact, if the manifold is
compact this is true and, if the manifold is not compact, the volume divergences can be easily
factorized. Thus
\beq
K_t(L_D)=\int dV_D K_t( L_D)(x)
\eeq
and
\beq
\ze(L_D,z)=\int dV_D \ze(L_D|z)(x),
\eeq
where $K_t( L_D)(x)$ and $\ze(L_D|z)(x)$ are the heat-kernel and the local
zeta-function, respectively.

However, if an internal symmetry is present, the scalar field is
vector valued, i.e. $\phi_i$ and the  simplest model is the $O(2)$
symmetry associated
with self-interacting charged fields in $R^4$. The Euclidean action is
\beq
S=\int  dx^4 \phi_i \aq \at-\lap+m^2 \ct \phi_i+\frac{\la}{4!}(\phi^2)^2
\cq
\:,\label{ea}\eeq
where $\phi^2=\phi_k \phi_k$ is the $O(2)$ invariant.
The Euclidean small disturbances operator
reads
\beq
A_{ij}=L_{ij}+\frac{\la}{6}\Phi^2 \de_{ik}+\frac{\la}{3}
\Phi_i \Phi_k,\qquad L_{ij}=\at -\lap+m^2
\ct\de_{ik}
\:,\label{sd}\eeq
in which $\Delta$ is the Laplace operator and
$\Phi$ the background field, assumed to be constant. Thus, one is actually
       dealing with a matrix-valued elliptic differential operator. In
this case,  the partition function is \cite{bens91-44-2480}
\beq
\ln Z=-\ln\det \left\| \frac{A_{ik}}{\mu^2}\right\| =-\ln\det \aq
\frac{(L+\frac{\la}{2}\Phi^2)}{\mu^2}\frac{(L+\frac{\la}{6}\Phi^2)}{\mu^2}\cq
\label{o2}
\eeq
As a consequence, one has to deal with the product of two elliptic
differential operators. In the case of a two-matrix, one has
\beq
\ln \det (AB) =\ln \det A+\ln \det B
\:.\label{ab}\eeq
Usually the way one proceeds is by
formally assuming the validity of the above
relation for differential operators. This may be quite ambiguous, since
one has to deal necessarily  with a regularization procedure.
 In fact, it turns out that the zeta-function
regularized determinants do {\it not} satisfy the above relation and, in
general,  there appears the so-called multiplicativity (or just
multiplicative) anomaly
\cite{kass89-177-199,kont95b}.
In terms of $F(A,B)\equiv \det (AB)/(\det A \det B)$ \cite{kont95b},
it is defined as:
\beq
a_D(A,B)=\ln F(A,B)=\ln \det (AB)-\ln \det (A)-\ln \det (B)
\:,\label{ma}\eeq
in which the determinants of the two elliptic operators, $A$ and $B$, are
assumed to be defined (e.g.,
regularized) by means of the zeta-function
\cite{ray71-7-145}.
It should be noted that the
non vanishing of the multiplicative anomaly implies that the relation
\beq
\ln \det A =\Tr \ln A
\label{tl}\eeq
does not hold, in general, for elliptic operators like $A=BC$.

It turns out that this  multiplicative anomaly
can be expressed by means of the non-commu\-ta\-ti\-ve residue
associated with a classical pseudo-differential operator, known as the
Wodzicki residue \cite{wodz87b}. Its important role in physics has
been recognized only recently. In fact, within the non-commutative
geometrical
approach to the standard model of the  electroweak interactions
\cite{conn90-18-29,conn96}, the Wodzicki residue is the {\it unique}
extension
of the Dixmier trace (necessary to write down the Yang-Mills action
functional) to the larger class of pseudo-differential operators ($\Psi$DO)
\cite{conn88-117-673}. Other recent contributions along these lines
are \cite{kala93u-31}-\cite{acke95u-6}. Furthermore, a proposal
to make use of the Wodzicki formulae as a practical tool in order
to determine the singularity
structure of zeta-functions has appeared in \cite{eliz96u-56} and the 
connection with the commutators anomalies of current algebras and the 
Wodzicki residue has been found in \cite{mick93}

The purpose of the present paper is to obtain explicitly the
multiplicative
anomaly for the product of two Laplace-like operators ---by direct
computations and by making use of several results due to Wodzicki--- and
to investigate the relevance of these concepts in physical situations.
As a result, the multiplicative anomaly will be
found to be vanishing for $D$ odd and also for $D=2$, being actually
present for
$D>2$, with $D$ even.

The contents of the paper are the following. In Sect. 2 we present some elementary computations
in order to show the highly non-trivial character of a
 brute force approach to the evaluation of the  multiplicative anomaly
associated with two  differential operators
(even with very simple ones). In Sect. 3 we briefly
recall several results due to Wodzicki, concerning the noncommutative
residue and a fundamental formula expressing the multiplicative anomaly
in terms of the corresponding residue of a suitable pseudo-differential
operator. In Sect.  4, the Wodzicki formula is used in the computation
of the multiplicative anomaly in $R^D$ and, as an example, the $O(2)$ model in
$R^4$ is investigated. In Sect. 5, a standard diagrammatic analysis of the
$O(2)$ model is discussed and evidence for  the  presence of
the multiplicative anomaly at this   diagrammatic level
is given. In Sect. 6 we treat the case of an arbitrary
compact smooth manifold without boundary.  Some final remarks are presented in the
Conclusions. In the Appendix a proof of the multiplicative anomaly formula
is outlined.

\section{ Direct calculations }

Motivated by the example discussed in the introduction, one might try
to perform a direct computation of the multiplicative anomaly in the case of
the two self-adjoint elliptic commuting operators
$L_p=-\lap+V_p$, $p=1,2$, in $ M_D$, with $V_p$ constant. Actually, we 
could deal with the shifts of two elliptic $\Psi$ODs. For the sake of
simplicity, we may put $\mu^2=1$ and consider all the quantities to be
dimensionless. At the end, one can easily restore $\mu^2$ by simple
dimensional considerations.

In order to
compute the multiplicative anomaly, one needs to obtain the
zeta-functions of the operators. Let us begin
with $M_D$ smooth and compact without boundary (the boundary case can 
be treated along the same lines) and let us try
to express $\zeta(s|L_1L_2)$ as a function of $\zeta(s|L_p)$.
If we denote  $L_0=-\lap$ and by $\la_i$ its non-negative,
discrete eigenvalues, the spectral theorem yields
\beq
\zeta(s|L_1L_2)=\sum_i \aq (\la_i+V_1) (\la_i+V_2) \cq^{-s}
\:.\label{l12}\eeq
Making use of the identity
\beq
(\la_i+V_1) (\la_i+V_2)= (\la_i+V_+)^2 -V_-^2   \:,
\label{e}
\eeq
with $V_+=(V_1+V_2)/2 $ and  $V_-=(V_1-V_2)/2 $, and noting that
\beq
\frac{V_{-}^2}{(\la_i+V_+)^2}< 1\,
\:,\label{e2}\eeq
for every individual $\la_i$, the binomial theorem gives
\beq
\aq (\la_i+V_1) (\la_i+V_2) \cq^{-s}=\sum_{k=0}^\ii
\frac{\Ga(s+k)}{k!\, \Ga(s)}
V_-^{2k}(\la_i+V_+)^{-2s-2k}
\:,\label{e3}\eeq
an absolutely convergent series expansion, valid without further
restriction.
Let us assume that $\Re s$ is large enough in order to safely commute the
sum over $i$ with the sum over  $k$. From the equations above, we get
\beq
\zeta(s|L_1L_2)=\zeta(2s|L_0+ V_+)+\sum_{k=1}^\ii
 \frac{\Ga(s+k)}{k! \, \Ga(s)}
V_-^{2k}\ze(2s+2k|L_0+V_+)
\:.\label{e4}\eeq
This series is convergent for large $\Re s$ and provides the
sought for analytical continuation to the whole complex plane.

To go further, we note that, when $ |c| < \la_1$ (smallest non-vanishing eigenvalue of 
$L$), one has
\beq
\zeta(s|L+c)=\zeta(s|L)+\sum_{k=1}^\ii
\frac{\Ga(s+k)}{k!\, \Ga(s)}
(-c)^{k}\ze(s+k|L)
\:,\label{e7}\eeq
Let us use this expression for $L_1$ and $L_2$. Since 
\beq
V_1=V_++V_-\,\,,V_2=V_+-V_- 
\:,\label{b}\eeq
one has
\beq
\zeta(s|L_1)=\zeta(s|L_0+V_++V_-)=\zeta(s|L_0+V_+)+\sum_{k=1}^\ii
\frac{\Ga(s+k)}{k!\, \Ga(s)}
(-V_-)^{k}\ze(s+k|L_0+V_+)
\:,\label{b1}\eeq
and
\beq
\zeta(s|L_2)=\zeta(s|L_0+V_+-V_-)=\zeta(s|L_0+V_+)+\sum_{k=1}^\ii
\frac{\Ga(s+k)}{k!\, \Ga(s)}
(V_-)^{k}\ze(s+k|L_0+V_+)
\:.\label{b2}\eeq
For $s=0$, there  are poles, but  adding the two zeta-functions for suitable 
$\Re s$ and making the separation between $k$ odd and $k$ even, all the 
terms associated with $k$ odd cancel. As a result 
\beq
\zeta(s|L_1)+\zeta(s|L_2)=2\zeta(s|L_0+V_+)+\sum_{m=1}^\ii
\frac{\Ga(s+2m)}{(2m)!\, \Ga(s)}
(V_-)^{2m}\ze(s+2m|L_0+V_+)
\:.\label{b3}\eeq
For suitable $\Re s$, from Eqs.~(\ref{e4}) and 
Eq.~(\ref{b3}) we may write
\beq
\zeta(s|L_1L_2)-\zeta(s|L_1)-\zeta(s|L_2)&=&\zeta(2s|L_0+V_+)-2\zeta(s|L_0+V_+) 
\nn \\
&+&\sum_{m=1}^\ii
\frac{(V_-)^{2m}}{\Ga(s)}\aq \frac{\Ga(s+m)}{m!}
\ze(2s+2m|L_0+V_+) \cp \nn \\
&-& \ap 2\frac{\Ga(s+2m)}{(2m)! }
\ze(s+2m|L_0+V_+)\cq
\:,\label{zz}\eeq

The multiplicative anomaly is minus the derivative with respect to $s$ in 
the limit $s \to 0$. Thus, it is present only when there are poles 
of the zeta functions evaluated at positive integer numbers bigger 
than $2$. From the 
Seeley theorem, the meromorphic structure of the zeta function related 
to an elliptic operator is known, also in manifolds with boundary, the 
residues at the poles being simply related to the Seeley-De Witt heat-kernel 
coefficients $A_r$. For example, For a D-dimensional manifold without boundary 
one has \cite{seeley}
\beq
\ze(z|L)=\frac{1}{\Ga(z)}\sum_{r=0}^\ii 
\frac{A_r}{z+r-\fr{D}{2}}+\frac{J(z)}{\Ga(z)}
\:,\label{st}\eeq 
$J(z)$ being the analytical part.
Since there are no poles at $s=0$ for $D$ odd and for $D=2$ in the zeta 
functions appearing on the r.h.s. of Eq.~(\ref{zz}),  
we can take the derivative at $s=0$, i.e.
\beq
a_D(L_1,L_2)=\sum_{m=1}^\ii(V_-)^{2m}\ze(2m|L_0+V_+) \aq
\frac{\Ga(m)}{\Ga(m+1)}-2\frac{\Ga(2m)}{\Ga(2m+1)} \cq
\:.\label{b33}\eeq
As a consequence, for $D$ odd and for $D=2$ the 
multiplicative anomaly is vanishing.

For $D >2$ and even, there are a finite number of simple poles other 
than at $s=0$ 
in Eq.~(\ref{zz}). As an example, in the important case $D=4$, in a compact manifold 
without boundary, the zeta function has simple poles at $s=2, s=1, s=0 
$, etc. Only the first one is relevant, the other being harmless. 
Separating the term corresponding to $l=1$, only this gives a non 
vanishing contribution when one takes the derivatives with respect to 
$s$ at zero. Thus, a direct computation yields
\beq
a_4(L_1,L_2)=\frac{A_0 V_-^2}{2}=\frac{\cal V_D}{4(4\pi)^2} \at V_1-V_2 
\ct^2
\:.\label{an4}\eeq
It 
follows that it exists potentially, an alternative direct method for computing the 
multiplicative anomaly for the shifts of two elliptic $\Psi$DOs and its structure will be a function of 
$V_-^2$ and of the  heat-kernel 
coefficients $A_r$, which, in principle, are computable (the first ones are 
known). We will come back on this point in Sect. 6, using the Wodzicki 
formula.

However, we observe that, here, the multiplicative anomaly is a function of the
series of zeta-functions related to operators of Laplace type.
One soon becomes convinced that it is not easy to go further along this way 
for an arbitrary D-dimensional manifold.

We conclude this section with explicit examples.

\medskip

\noindent{\sf Example 1:} $M_D=R^D$.

Let us start with  a particularly simple example, i.e.
$M_D=R^D$.
 The two zeta-functions
$\zeta(s|L_i)$ are easy to evaluate and read
\beq
\zeta(s|L_i)=\frac{\ca V_D}{(4\pi)^{\fr{D}{2}}}V_i^{\fr{D}{2}-s}
\frac{\Ga(s-\fr{D}{2})}{\Ga(s)},
 \qquad i=1,2 \:, \label{xxx1}\eeq
where $\ca V_D$ is the (infinite) volume of $R^D$. We need to compute
$\zeta(s|L_1L_2)$. For $\Re s > D/2$, starting from the
 spectral definition,
one gets
\beq
\zeta(s|L_1L_2)=\frac{2\ca V_D}{4\pi)^{\fr{D}{2}} \Ga(\fr{D}{2})}
\int_0^\ii dk k^{D-1} \aq k^4+(V_1+V_2)k^2+V_1V_2 \cq^{-s}
\:.\label{lu}\eeq
For $\Re s > (D-1)/4$, the the above integral can be evaluated
\cite{grad80b},
to yield
\beq
\zeta(s|L_1L_2)=\frac{\sqrt 2\pi \ca V_D \Ga(2s-\fr{D}{2})}{2^s (4\pi)^{\fr{D}{2}}
\Ga(s)}
\at  \al^2-1 \ct^{\fr{1-2s}{4}} \at V_1V_2\ct^{\fr{D}{4}-s}
P^{{\fr{1}{2}-s}}_{{s-\fr{D+1}{2}}}(\al)
\:,\label{z12}\eeq
 $P^\mu_\nu(z)$ being the associate Legendre function of the first
kind (see for example \cite{grad80b}), and
\beq
\al=\frac{V_1+V_2}{2 \sqrt {V_1V_2}}
\:.\label{al}\eeq
This provides the analytical continuation to the whole complex plane.
For $D=2Q+1$,  one easily gets
\beq
\zeta(0|L_1L_2)&=&0, \nn \\
\zeta'(0|L_1L_2)&=&\frac{\sqrt 2\pi \ca V_D \Ga(-Q-\fr{1}{2})}{ (4\pi)^{\fr{D}{2}}}
\at  \al^2-1 \ct^{\fr{1}{4}} \at V_1V_2\ct^{\fr{D}{4}}
P^{{\fr{1}{2}}}_{{-\fr{D+1}{2}}}(\al) \nn \\
&=&\frac{ \ca V_D \Ga(-Q-\fr{1}{2})}{ (4\pi)^{\fr{D}{2}}}
\aq 2 (V_1V_2)^{\fr{D}{2}} (1+\cosh (D\ga)) \cq^{1/2}
\:,\label{zp}\eeq
in which $\cosh \ga=\al$. The first equation says that the conformal
anomaly  vanishes.
On the other hand, one has for $D$ odd
\beq
\zeta'(0|L_1)+\zeta'(0|L_2)=\frac{ \ca V_D \Ga(-Q-\fr{1}{2})}{
(4\pi)^{\fr{D}{2}}}
\at V_1^{\fr{D}{2}}+ V_2^{\fr{D}{2}} \ct
\:,\label{9}\eeq
As a consequence, making use of elementary properties of the
hyperbolic cosine, one gets $a(L_1,L_2)=0$. Namely, for $D$ odd the
multiplicative anomaly is vanishing (see \cite{kont95b}).

For $D=2Q$, the situation is much more complex. First the
conformal anomaly is non-zero, i.e.
\beq
\zeta(0|L_1L_2)=\frac{ \ca V_D}{ (4\pi)^{Q}}\frac{(-1)^Q}{Q!}
\aq  (V_1V_2)^{Q/2} \cosh (Q\ga) \cq
\:,\label{10}\eeq
and, in general, the multiplicative anomaly is present.
As a check, for $D=2$, we get
\beq
\zeta(0|L_1L_2)=-\frac{ \ca V_2}{ 4\pi}
\aq  (V_1V_2)^{1/2} \cosh \ga \cq=-\frac{ \ca V_2}{
4\pi}(V_1+V_2)=\frac{1}{ 4\pi}a_1(A)=\ze(0|A)
\:,\label{11}\eeq
where $A=-\lap I+V$ is a $2 \times 2$ matrix-valued differential
operator, $ I$  the identity matrix, $ V=$ diag $(V_1,V_2)$,
 and $a_1(A)$ is the first related
Seeley-De Witt coefficient, given by the well known expression $\int dx^2
(-\tr V)$.

Unfortunately, it is not simple to write down ---within this naive
approach--- a reasonably simple expression for it, because  the
associate
Legendre function depends on $s$ through the two indices $\mu$ and
$\nu$. However, it is easy to show that the anomaly is absent
when $V_1=V_2$, therefore it will depend only on the difference
$V_1-V_2$. Thus, one may consider the case $V_2=0$. As a result,
Eq.~(\ref{lu}) yields the  simpler expression
\beq
\zeta(s|L_1L_2)&=&\frac{\sqrt 2\pi \ca V_D }{
(4\pi)^{\fr{D}{2}}\Ga(\fr{D}{2})}
\frac{\Ga(\fr{D}{2}-s)\Ga(2s-\fr{D}{2})}{\Ga(s)}
 V_1^{\fr{D}{2}-2s}
\:.\label{xxx2}\eeq
In this case the multiplicative anomaly is given by
\beq
a(L_1,L_2)=\ln \det (L_1L_2)-\ln \det (L_1)
\:,\label{mar}\eeq
since the regularized quantity $\ln \det (L_2)=0$.
It is easy to show that, when $D$ is odd, again $
a_D(L_1,L_2)=0$. When $D=2Q$, one obtains
\beq
a_{2Q}(L_1,L_2)=\frac{ \ca V_D}{ (4\pi)^{Q}}\frac{(-1)^Q}{2Q!}
  V_1^{Q} \aq \Psi(1)-\Psi(Q) \cq
\:.\label{anom}\eeq
We conclude this first
example by observing that the multiplicative anomaly is
absent when $Q=1$,  $D=2$, and that it
is present for $ Q>1$, $D >2$ even. The
result obtained is partial and more powerful techniques are necessary in
order to deal with the general case. Such techniques will be
introduced in the next section.
\medskip

\noindent{\sf Example 2:} $M_D=S^1\times R^{D-1}$,
$D=1,2,3,\ldots$

In this case the zeta functions corresponding to $L_i$, $i=1,2$,
 are given by
\beq
\zeta(s|L_i) = \frac{\pi^{(D-1)/2-2s}
\Gamma(s+(1-D)/2)}{2^{2s+1} L^{D-2s}\Gamma(s)}
\sum_{n=-\infty}^{\infty} \left[n^2 + \left( \frac{L}{2\pi}
\right)^2V_i \right]^{(D-1)/2-s}
\eeq
($i=1,2$, here $L$ is the length of $S^1$).
In terms of the basic zeta function (see \cite{eliz94-35-6100}):
\begin{eqnarray}
 \zeta(s;q)& \equiv & \sum_{n=-\infty}^{\infty} (n^2+q)^{-s}
 \\ && =
\sqrt{\pi} \, \frac{\Gamma(s-1/2)}{\Gamma(s)} q^{1/2-s} +
\frac{4\pi^s}{\Gamma(s)} q^{1/4-s/2} \sum_{n=1}^\infty n^{s-1/2}
K_{s-1/2} (2\pi n \sqrt{q}), \nonumber
\end{eqnarray}
where $K_\nu$ is the modified Bessel function of the second kind,
we obtain
\begin{eqnarray}
\zeta(s|L_i) &=& \frac{\pi^{-D/2}}{
 \Gamma(s)} \left[ 2^{-D}L \Gamma(s-D/2)
  V_i^{D/2-s} \right. \nonumber \\
&& \left. + 2^{2-s-D/2} L^{s+1-D/2}
 V_i^{D/4-s/2}
\sum_{n=1}^{\infty} n^{s-D/2} K_{s-D/2} (n L \sqrt{V_1})\right] \\
&& \equiv \zeta^{(1)} (s|L_i) + \zeta^{(2)} (s|L_i).  \nonumber
\end{eqnarray}
For the determinant we get, for $D$ odd,
\begin{eqnarray}
\det L_i &=&  \exp \left\{ -  \pi^{-D/2} \left[ 2^{-D} L
\Gamma(-D/2)
V_i^{D/2} \right. \right. \nonumber \\ && \left. \left.+ (2L)^{1-D/2}
V_i^{D/4} \sum_{n=1}^{\infty} n^{-D/2} K_{D/2}
 (n L \sqrt{V_i})\right] \right\},
\end{eqnarray}
 for $D$ even ($D=2Q$),
\begin{eqnarray}
\det L_i &=& \exp \left[ -\frac{L}{Q!} \left(
-\frac{1}{4\pi}\right)^Q V_i^Q \left( \sum_{j=1}^Q \frac{1}{j} -
\ln V_i \right) \right. \nonumber \\ && +  \left.
 4L \left( \frac{\sqrt{V_i}}{2\pi L} \right)^Q
 \sum_{n=1}^{\infty} n^{-Q} K_Q (n L \sqrt{V_i})\right] .
\end{eqnarray}
As for the product $L_1L_2$, using the same strategy as before,
after some calculations we obtain (here we use the short-hand notation
$L_\pm \equiv L_0 + V_\pm$, cf. equations above):
\begin{eqnarray}
\det (L_1L_2) &=& (\det L_+)^2 \exp  \left\{ -\sum_{p=1}^{[Q/2]} 2L
\frac{V_{-}^{2p}(-V_+)^{Q-2p}}{(2p)!(Q-2p)!(4\pi)^Q} \right. \nonumber \\
&&  \times \left[ 1-C + \frac{1}{2 \ (Q-2p)!} +
\frac{1}{2}  \sum_{j=1}^{p-1} \frac{1}{j} - \psi (2p) - \ln V_+ \right]
\nonumber \\ && - \left. \sum_{p=[Q/2]+1}^\infty \frac{V_{-}^{2p}}{p}
\zeta^{(1)} (2p| L_+) - \sum_{p=1}^\infty \frac{V_{-}^{2p}}{p}
\zeta^{(2)} (2p| L_+) \right\},
\end{eqnarray}
where $[x]$ means `integer part of $x$' and  $C$ is the Euler-Mascheroni
 constant.
We can check from these formulas    that the anomaly
(\ref{ma}) is zero in the case of odd dimension $D$. Actually, this is
most easily seen, as before, by using the expression corresponding to
(\ref{lu}) for the present case. It also vanishes for $D=2$.
The formula above is useful in
order to obtain numerical values for the case $D$ even, corresponding
to different values of $D$ and $L$ (the series converge very quickly).
The results are given in Table 1. We have looked at the variation of the
anomaly in terms of the different parameters: $L, D, V_1$ and $V_2$
while keeping the rest of them fixed. Within numerical errors, we
have checked the complete coincidence with formula (\ref{wod55}) in   Sect. 4.

\begin{table}

\begin{center}

\begin{tabular}{|c|c|c|c||c|}
\hline \hline
$L$ & $D$ & $V_1$ & $V_2$ & $a(L_1,L_2)$ \\
 \hline \hline
 1 & 2 & 2 & 2 & 0.
 \\ \hline
0.1 & 2 & 8 & 3 & --$1.8686 \times 10^{-14}$
 \\ \hline
1 & 2 & 8 & 3 & --$2.0817 \times 10^{-17}$
 \\ \hline
5 & 2 & 8 & 3 & --$1.4572 \times 10^{-16}$
 \\ \hline
10 & 2 & 8 & 3 & --$1.4572 \times 10^{-16}$
 \\ \hline
1 & 2 & 10 & 1 & $2.87 \times 10^{-12}$
 \\ \hline
1 & 4 & 10 & 1 & 0.064117
 \\ \hline
1 & 6 & 10 & 1 & --0.028063
 \\ \hline
1 & 8 & 10 & 1 & 0.0151245
 \\ \hline
1 & 10 & 10 & 1 & --0.003636
 \\ \hline
1 & 12 & 10 & 1 & 0.0006124
 \\ \hline
1 & 14 & 10 & 1 & --0.00008166
 \\ \hline
1 & 16 & 10 & 1 & $9.09 \times 10^{-6}$
 \\ \hline
 1 & 4 & 2 & 1 & 0.0007916
 \\ \hline
 1 & 4 & 5 & 2 & 0.007124
 \\ \hline
 1 & 4 & 1 & 6 & 0.019789
 \\ \hline
 1 & 6 & 2 & 1 & --0.0000945
 \\ \hline
 1 & 6 & 5 & 2 & --0.001984
 \\ \hline
 1 & 6 & 1 & 6 & --0.005512
 \\ \hline
 0.1 & 4 & 7 & 2 & 0.001979
 \\ \hline
 0.5 & 4 & 7 & 2 & 0.009895
 \\ \hline
 1 & 4 & 7 & 2 &  0.019789
 \\ \hline
 2 & 4 & 7 & 2 &  0.0395786
 \\ \hline
 5 & 4 & 7 & 2 &  0.098947
 \\ \hline
 10 & 4 & 7 & 2 & 0.197893
 \\ \hline
 20 & 4 & 7 & 2 & 0.395786
 \\ \hline
 0.1 & 6 & 7 & 2 & --0.00070865
 \\ \hline
 0.5 & 6 & 7 & 2 & --0.00354326
 \\ \hline
 1 & 6 & 7 & 2 &  --0.0070865
 \\ \hline
 2 & 6 & 7 & 2 &  --0.014173
 \\ \hline
 5 & 6 & 7 & 2 &  --0.0354326
 \\ \hline
 10 & 6 & 7 & 2 & --0.07008652
 \\ \hline
 20 & 6 & 7 & 2 & --0.141730
\\ \hline \hline \end{tabular}

\caption{{\protect\small Values of the multiplicative anomaly
$a(L_1,L_2)$
in terms of the parameters: $L, D, V_1$ and $V_2$. Observe its evolution
when some of the parameters are kept fixed while the others are varied.
In all cases, a perfect coincidence with Wodzicki's expression for the
anomaly is obtained (within numerical errors).}}

\end{center}

\end{table}

\medskip

\noindent{\sf Example 3:} $M_D= R^D$ {\sf with Dirichlet b.c. on
$p$ pairs of perpendicular hyperplanes}.

The zeta function is, in this case,
\beq
\zeta(s|L_i) = \frac{\pi^{(D-p)/2-2s}
\Gamma(s+(p-D)/2)}{2^{D-p+1} \prod_{j=1}^p a_j \, \Gamma(s)}
\sum_{n_1,\ldots,n_p=1}^{\infty} \left[\sum_{j=1}^p \left(
\frac{n_j}{a_j}\right)^2 + V_i \right]^{(D-p)/2-s},
\eeq
where the  $a_j$, $j=1,2,\ldots, p$,
 are the pairwise separations between the perpendicular hyperplanes.
For the determinant, we get, for $D-p=2h+1$ odd,
\beq
\det L_i =  \exp \left\{ -
\frac{\pi^{h+1/2}}{2^{2h+2} \prod_{j=1}^p a_j} \Gamma
(-h-1/2) \sum_{n_1,\ldots,n_p=1}^{\infty} \left[\sum_{j=1}^p
 \left( \frac{n_j}{a_j}\right)^2 + V_i \right]^{h-1/2}\right\},
\eeq
and, for $D-p=2h$ even,
\begin{eqnarray}
\det L_i &=& \exp \left\{ \frac{(-\pi)^{h}}{2^{2h+1} h!\,
\prod_{j=1}^p a_j} \left[ \left( 2+ h \sum_{j=1}^{h-1} \frac{1}{j}
\right)\sum_{n_1,\ldots,n_p=1}^{\infty} \left[\sum_{j=1}^p
 \left( \frac{n_j}{a_j}\right)^2 + V_i \right]^h \right.\right.
\nonumber \\ && \left. \left.  +
\sum_{n_1,\ldots,n_p=1}^{\infty} \left[\sum_{j=1}^p
 \left( \frac{n_j}{a_j}\right)^2 + V_i \right]^h
\ln \left[\sum_{j=1}^p
 \left( \frac{n_j}{a_j}\right)^2 + V_i \right]
\right] \right\}.
\end{eqnarray}
For the calculation of the anomaly one follows the same steps of the two
preceding examples and we are not going to repeat this again.
In order to obtain the final numbers one must make use of the
inversion formula for the Epstein zeta functions of these
expressions \cite{eliz94-35-6100,eliz94b}.


\section{The Wodzicki residue and the multiplicative anomaly}

For reader's convenience, we will review in this section the necessary
information concerning the Wodzicki residue  \cite{wodz87b} (see,
also \cite{kass89-177-199} and the references to Wodzicki
 quoted therein)
that will be used in the rest of the paper.
Let us consider a D-dimensional smooth compact manifold without boundary $M_D$
and a (classical) $\Psi$DO, $A$, of order
$m$,
 acting on sections of vector
bundles on $M_D$. To any $\Psi$DO, $A$, it corresponds a complete
symbol $a(x,k)$, such that, modulo infinitely
smoothing operators, one has
\beq
(A f)(x)\sim \int_{R^D}\frac{dk}{(2\pi)^{D}}\int_{R^D}dy
e^{i(x-y)k}a(x,k)f(y)
\:.\label{sy}\eeq
The complete symbol admits an asymptotic expansion for $|k| \to \ii$,
given by
\beq
a(x,k)\sim \sum_j a_{m-j}(x,k)
\:,\label{sy1}\eeq
and fulfills the homogeneity property $ a_{m-j}(x,tk)= t^{m-j}a_{m-j}(x,k)$,
for $t>0$. The number $m$ is
called the order of $A$.

If $P$ is an elliptic operator of order $p>m$, according to
Wodzicki one has the following property of the non-commutative residue, which
we may take as its characterization. \medskip

\noindent {\bf Proposition.} The trace of the operator  $AP^{-s}$ exists
and admits a meromorphic continuation to the whole complex plane, with a
simple pole at $s=0$. Its Cauchy residue  at $s=0$ is proportional
to the so-called non-commutative (or Wodzicki) residue of $A$:
\beq
\mbox{res} (A)=p \Res_{s=0} \Tr (AP^{-s})
\:.\label{wod1}\eeq
The r.h.s. of the above equation  does not depend on $P$ and is taken as
the definition of the
Wodzicki residue of the $\Psi$DO,  $A$.      \medskip

\noindent {\bf Properties.}
(i) Strictly related to the latter
result is the one which follows, involving the short-$t$ asymptotic expansion
\beq
\Tr (A e^{-tP})\simeq \sum_j \al_j t^{\fr{D-j}{p}-1}-\frac{\mbox{res}
(A)}{p} \ln t+O(t \ln t)
\:.\label{wod11}\eeq
Thus, the Wodzicki residue of $A$, a $\Psi$DO, can be read off
from the above asymptotic expansion selecting the coefficient
proportional to $\ln t$.

(ii) Furthermore, it is possible to show that  $\mbox{res} (A)$ is
linear with respect to $A$ and possesses the  important
property of being the unique trace on the algebra of
the $\Psi$DOs, namely, one has $\mbox{res}(AB)=\mbox{res}(BA)$. This last
property has deep implications when including gravity within
 the non-commutative geometrical approach
to  the Connes-Lott model of the electro-weak interaction theory
\cite{conn88-117-673,conn90-18-29,conn96}.

(iii) Wodzicki has also obtained a local form of the
non-commutative residue, which has the fundamental consequence of
characterizing it through a scalar density. This density       can be
integrated to yield the Wodzicki residue, namely
\beq
\mbox{res}(A)=\int_{M_D}\frac{dx}{(2\pi)^{D}}\int_{|k|=1}a_{-D}(x,k)dk
\:.\label{wod2}\eeq
Here the component of order $-D$ of the complete symbol appears.
Form the above result it immediately follows that $\mbox{res} (A)=0$
when $A$ is an elliptic  differential operator.

(iv) We conclude this summary with the multiplicative anomaly formula,
again due to Wodzicki.  A more general formula has
been derived in \cite{kont95b}.
Let us consider two invertible elliptic self-adjont
operators, $A$ and $B$, on $M_D$. If we  assume that they commute, then the
following formula applies
\beq
a(A,B)=\frac{\mbox{res}\aq (\ln(A^bB^{-a}))^2
\cq}{2ab(a+b)}=a(B,A)
\:,\label{wod3}\eeq
where $a >0$ and $ b> 0$ are the orders of $A$ and $B$, respectively.
A sketch of the proof is presented in the Appendix.
It should be noted that $a(A,B)$ depends on a  $\Psi$DO  of zero
order. As a consequence, it is independent on the renormalization
scale $\mu$ appearing in the path integral.

(v) Furthermore, it can be
iterated consistently. For example
\beq
\ze'(A,B)&=&\ze'(A)+\ze'(B)+a(A,B),\label{wod4} \\
\ze'(A,B,C)&=&\ze'(AB)+\ze'(C)+a(AB,C)=\ze'(A)+\ze'(B)+\ze'(C)+a(A,B)+a(AB,C)
\:.\nn \eeq
As a consequence,
\beq
a(A,B,C)=a(AB,C)+a(A,B)
\:.\label{wod5}\eeq
Since $a(A,B,C)=a(C,B,A)$, we easily obtain the cocycle condition (see
\cite{kont95b}):
\beq
a(AB,C)+a(A,B)=a(CB,A)+a(C,B)
\:.\label{wod6}\eeq

\section{The $O(2)$ bosonic model}

In this section we come back to the problem of the exact computation of the
  multiplicative anomaly
in the model considered in Sect.  2. Strictly speaking, the result of
the last section is valid for a compact manifold, but in the case of $R^D$
the divergence is trivial, being contained in the volume factor.
The Wodzicki formula gives
\beq
a(L_1,L_2)=\frac{1}{8} \, \mbox{res}\aq (\ln(L_1 L_2^{-1}))^2 \cq
\:.\label{wod44}\eeq
We have to construct the complete symbol of the $\Psi$DO of zero order
 $[\ln (L_1 L_2^{-1})]^2$.
It is given by
\beq
a(x,k)=\aq \ln (k^2+V_1)-\ln (k^2+V_2) \cq^2
\:.\label{90}\eeq
For large $k^2$, we have the following expansion, from which one can easily
 read off the  homogeneuos components:
\beq
a(x,k)= \sum_{j=2}^\ii c_j k^{-2j}= \sum_{j=2}^\ii a_{2j}(x,k)
\:,\label{91}\eeq
where
\beq
c_j=\sum_{n=1}^j\frac{ (-1)^j}{n(j-n)} \at V_1^n-V_2^n \ct \at
V_1^{j-n}-V_2^{j-n} \ct
\:.\label{92}\eeq
As a consequence, due to the local formula one immediately gets the
following result: for $D$ odd, the multiplicative anomaly
vanishes, in perfect agreement with the direct calculation of Sect.  2. This
result is consistent with a general theorem contained in
\cite{kont95b}.

For $D$ even, if $D=2$ one has no multiplicative anomaly, 
while for $D=2Q$, $Q>1$, one gets
\beq
a(L_1,L_2)=\frac{\ca V_D (-1)^Q}{4(4\pi)^Q \Ga(Q)}\sum_{j=1}^{Q-1}
\frac{1}{j(Q-j)} \at V_1^j-V_2^j \ct \at
V_1^{Q-j}-V_2^{Q-j} \ct
\:.\label{wod55}\eeq
It is easy to show that for $V_2=0$ this expression reduces to the one
obtained directly in Sect. 2.

In the  $O(2)$ model, for $D=4$, we have
\beq
a(L_1,L_2)=\frac{\ca V_4 }{4(4\pi)^2}\at V_1-V_2 \ct^2=\frac{\ca V_4
}{36(4\pi)^2}\la^2 \Phi^4
\:,\label{6}\eeq
which, for dimensional reasons,  is independent of the renormalization
parameter $\mu$. Then, the one-loop effective potential reads
\beq
V_{eff}=-\frac{\ln Z}{\ca V_4}=\frac{M_1^4}{64\pi^2}\at-\frac{3}{2}+\ln
\frac{M_1^2}{\mu^2} \ct+\frac{M_2^4}{64\pi^2}\at-\frac{3}{2}+\ln
\frac{M_2^2}{\mu^2} \ct+\frac{1}{72(4\pi)^2}\la^2 \Phi^4
\:,\label{ep}\eeq
with
\beq
M_1^2=m^2+\frac{\la}{2}\Phi^2,  \qquad
M_2^2=m^2+\frac{\la}{6}\Phi^2
\:.\label{7}\eeq
Thus, the additional multiplicative
anomaly contribution seems to modify the usual Coleman-Weinberg potential. A more
careful analysis is required in order to investigate the consequences
of this remarkable fact.

\section{Feynmann diagrams}

The necessity of the presence of the multiplicative anomaly in quantum
field theory can also be understood perturbatively,
using the background field method. The effective action of the $O(2)$
model in a background field $\Phi$ will be denoted by $\Ga(\Phi,\phi)$,
where $\phi$ is the mean field. Then, if $\Ga_0(\phi)$ denotes the
effective action with vanishing $\Phi$, it turns out that
\beq
\Ga(\Phi,\phi)=\Ga_0(\Phi+\phi).
\eeq
Therefore, the $n$-th order derivatives of $\Ga$ with respect to $\phi$
at $\phi=0$ determine the vertex functions of the $O(2)$ model in the
background external field. The one-loop approximation to $\Ga$ is
again given by $\log\det(L_1L_2)$, and the determinant of either of the
operators, $L_1$ and $L_2$, corresponds to the
sum of all vacuum-vacuum $1$PI  diagrams where only particles of masses
squared $M_1^2=m^2+\la\Phi^2/2$ or $M_2^2=m^2+\la\Phi^2/6$ flow along
the internal lines. In Fig. 1 we have depicted this, by using a solid line
for type-$1$ particles and a dashed line for type-$2$ particles.

\begin{center}
\leavevmode
\epsfysize=5cm
\epsfbox{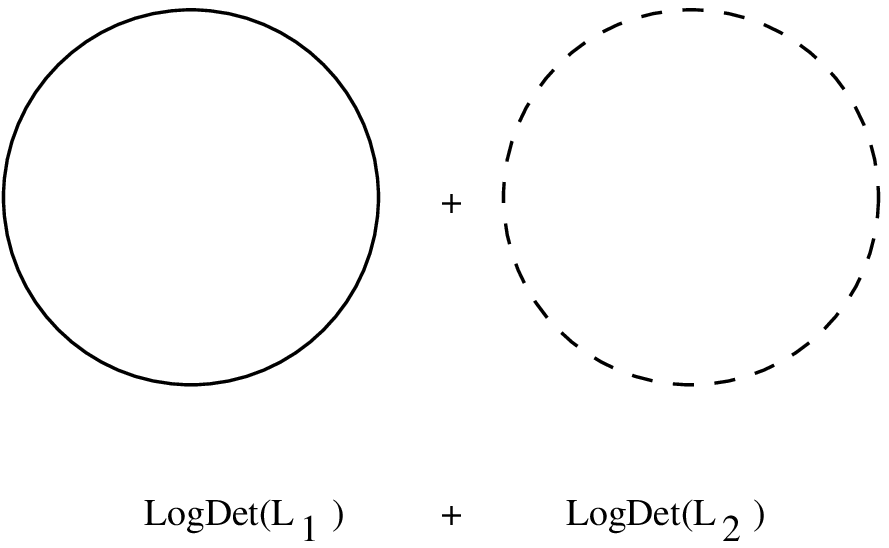}
\end{center}

\bigskip

{\small Figure 1. The Feynmann graph giving the one-loop effective
potential without taking into account the anomaly.}

\bigskip

Thus, for example, the inverse propagator at zero momentum for type-$1$ 
particle, as computed from the above effective potential, is
obtained from the second derivative with respect to $\phi_1$. The only
$1$PI graphs which contribute are shown in Fig. 2.

\begin{center}
\leavevmode
\epsfxsize=13cm
\epsfbox{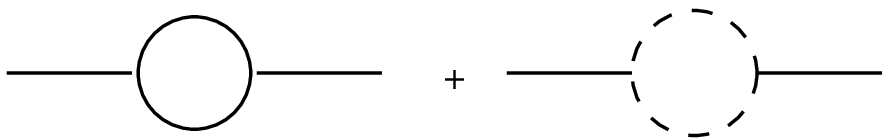}
\end{center}

\bigskip

{\small Figure 2.     Contributions coming from 1PI graphs.}

\bigskip

This is clearly not the case, as the full theory exhibits a trilinear
coupling $\phi_2(\phi_1)^2$ which gives the additional Feynmann graph
depicted in Fig. 3.

\begin{center}
\leavevmode
\epsfysize=2cm
\epsfbox{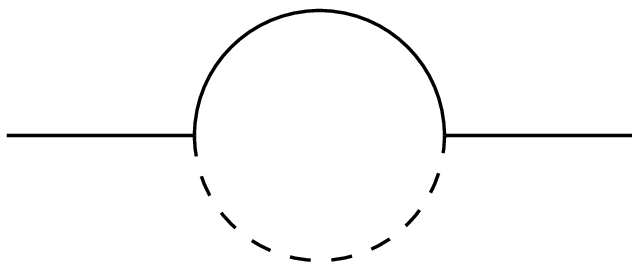}
\end{center}

\bigskip

{\small Figure 3. Additional Feynman graph of the full theory.}

\bigskip

Without investigating this question any further, we can safely affirm
already
that a perturbative formula for the Wodzicki anomaly given in terms of
Feynmann diagrams should exist. It surely owes its simple form to  very
subtle cancellations among an infinite class of Feynmann diagrams.

We conclude this section with some remarks.
In the present model, the existence of a
 multiplicative anomaly of the type considered could be
a trivial problem, in fact it has the same form as the
classical potential energy. This suggests that it can be absorbed in a
finite
renormalization of the coupling constant of the theory. Secondly, this
anomaly
gives no contribution to the one-loop beta function of the model, since
it is independent of the arbitrary renormalization scale, but it certainly
contributes to the two-loop beta function. And, finally, we have seen
that the anomaly can be interpreted as an external field effect which,
 in the present model,
could be relevant only when the theory is coupled to an external
source. Therefore, it should be very interesting to study its
relevance in at least two other situations, namely the cases
 of a spontaneously
broken symmetry and of QED in external background fields.

\section{The case of a general, smooth and compact manifold $M_D$
without boundary}

Since the multiplicative anomaly is a local functional, it is
possible to express it in terms of the Seeley-De Witt spectral coefficients. Let
us consider again the operator $L_p=L_0+V_p$, with $L_0=-\lap$ acting on
scalars, in a smooth and compact manifold $M_D$
without boundary. We have to compute the Wodzicki residue of the
$\Psi$DO
\beq
\aq \ln (L_1L_2^{-1})\cq^2
\:.\label{z1}\eeq
With this aim, if $V_1 < V_2$, we can consider the $\Psi$DO
\beq
\aq \ln (L_1L_2^{-1})\cq^2 e^{-tL_1}
\:,\label{z2}\eeq
and compute the $\ln t$ term in the short-$t$ asymptotic expansion of its trace.
We are dealing here with self-adjoint operators and thus, by using the spectral
 theorem, we get
\beq
\Tr \aq \at \ln (L_1L_2^{-1})\ct^2 e^{-t L_1} \cq=
\int_{V_1}^\ii d\la \rho(\la|L_1)\aq \ln \la- \ln (\la +V_2-V_1) \cq^2
e^{-t \la}
\:,\label{z3}\eeq
where $\rho(\la|L_1)$ is the spectral density of the self-adjoint
operator $L_1$.

Now, it is well known that the short-$t$ expansion of the above trace receives
contributions from the asymptotics, for large $\la$, of the integrand in the spectral integral.
The asymptotics of the spectral function associated with $L_1$ are
known to be given by (see, for example
\cite{horm68-121-193,cogn89-223-416}, and the references
therein)
\beq
\rho(\la|L_1)\simeq \sum_{r=0}^{r <D/2} \frac{A_r(L_1)}{\Ga(\fr{D}{2} -r)}
\la^{\fr{D}{2}-r-1}
\:,\label{z4}\eeq
here the quantities $A_r(L_1)$ are the Seeley-De Witt heat-kernel coefficients 
while, for large $\la$, we have in addition
\beq
\aq \ln \la- \ln (\la+V_2-V_1) \cq^2 \simeq \sum_{j=2}^\ii b_j \la^{-j}
\:,\label{z5}\eeq
being the $b_j$ computable, for instance
$b_2=(V_2-V_1)^2\,,\,b_3=-2(V_2-V_1)^3$, etc.
As a result, we get the short-$t$ asymptotics in the form
\beq
\Tr \aq \at \ln L_1L_2^{-1}\ct^2 e^{-tL_1} \cq\simeq \sum_{r=0}^{r <D/2}
\frac{A_r(L_1)}{\Ga(\fr{D}{2}
-r)} \sum_{j=2}^\ii b_j t^{r+j-\fr{D}{2}}\Ga(\fr{D}{2}-r-j,tV_1)
\:,\label{z6}\eeq
where $\Ga(z,x) $  the incomplete gamma function. From this expression 
one obtains the following results:

(i) If $D$ is odd, say $D=2Q+1$, the first argument of the incomplete
gamma function is never zero or a negative integer. Thus, the $\ln t$
is absent and, from the Wodzicki theorem, the multiplicative anomaly is
absent too, again in agreement with the Kontsevich-Vishik theorem
\cite{kont95b} and the explicit calculations in the previous sections.

(ii) If $D$ is even, we have to search for the log terms only, that is
$ -Q+r+j=0$, for $r \geq 0$ and $j \geq 2$. As a result, for $D=2$
 the log term is absent once more,  again in agreement with  the explicit
calculations of the previous sections. The multiplicative anomaly  is present
starting from  $D \geq 4$. In the important case when $D=4$, it turns
out that the multiplicative anomaly  is identical to the one,
related with $R^4$, that has been evaluated previosly.
Terms depending on the
curvature become operative only for $D \geq 6$.

\section{Conclusions}

In this paper, the multiplicative anomaly associated with the
zeta-function regularised determinant of two $\Psi$DOs of
Laplace type on a D-dimensional smooth manifold without boundary has been
studied. From a physical point of view, this condition does not seem to
be too
restrictive, because the one-loop effective potential
may be expressed as a logarithm of the determinant of such kind of
elliptic differential operators.

We have shown how a direct calculation leads to analytical
difficulties, even in the most simple examples. Fortunately, a very elegant
formula for the multiplicative anomaly has been found by Wodzicki
and we have used it here in order
 to compute the anomaly explicitly. It is worth
 mentioning
that, from a computational point of view, this constitutes
 a big improvement,
since one can make use of the results
concerning the computation of one-loop
effective potential, related to second order elliptic differential 
operators of Laplace type. Furthermore, within the background
field method, we have identified
the presence of the multiplicative anomaly in the
diagrammatic perturbative approach too.

With regard to our example, namely the product $L_1L_2$,
 we have shown that the multiplicative anomaly
is vanishing for $D$ odd and also for $D=2$. This seems to be related with
the fact that we have only considered differential operators
 of second order
(Laplace type). For first-order differential operators
(Dirac like),  things could be quite different, in principle,
and we will consider this important case elsewhere.

Another interesting issue is the generalization of all these procedures
 to smooth manifolds with a boundary. Again one should expect
to obtain different results in those situations.

\ack{We would like to thank Guido Cognola and  Klaus Kirsten for valuable
discussions. This work has been supported by the cooperative agreement
INFN (Italy)--DGICYT (Spain).
EE has been partly financed by DGICYT (Spain), project PB93-0035, and
by  CIRIT (Generalitat de Catalunya),  grant 1995SGR-00602.
}

\appendix

\section{Appendix:
         The Wodzicki formula for the multiplicative anomaly}

In this Appendix, for the reader's convenience we  present a proof  of
the multiplicative anomaly  formula along the lines of Ref.
\cite{kont95b}.

Recall that if $P$ is an elliptic operator of order $p>a$, according to
Wodzicki, one has the following property of the non-commutative residue
related to the $\Psi$DO $A$: in a neighborhood of $z=0$, it holds
\beq
z\Tr (AP^{-z})=\frac{1}{ \Ga(1+z)} \frac{\mbox{res} (A)}{p}+O(z^2)
\:.\label{A1}\eeq
Now we resort to the following
\medskip

\noindent{\bf Lemma}. If $\eta$ is a $\Psi$DO of zero order, $a$, and
$B$ a $\Psi$DO
of positive order, $b$, and $\ga$ and $x$  positive real numbers
 then, in a neighborhood of
$s=0$, one has
\beq
s\Tr (\ln \eta \eta^{-xs}B^{-\ga s})=\frac{\mbox{res} (\ln \eta)}{
\Ga(1+\ga s) \ga b}- sx \frac{\mbox{res} ((\ln \eta)^2)}{
\Ga(1+\ga s) \ga b}+O(s^2)
\:.\label{a2}\eeq
The Lemma is a direct consequence of the formal expansion
\beq
\eta^{-xs}=e^{-xs\ln \eta}=I-xs \ln \eta +O(s^2)
\:\label{vl}\eeq
and of Eq.~(\ref{A1}). From the above Lemma, it  follows that
\beq
\lim_{s \to 0}  \partial_s \aq s\Tr (\ln \eta \eta^{-xs}B^{-\ga
s})\cq  =
C \frac{\mbox{res} (\ln \eta)}{b}- x \frac{\mbox{res} [(\ln \eta)^2]}{\ga b}
\:,\label{a3}\eeq
in which $C$ is the Euler-Mascheroni constant.

Now consider two invertible, commuting, elliptic, self-adjont
operators $A$ and $B$ on $M_D$, with $a$ and $b$ being the orders
 of $A$ and $B$, respectively.
Within the zeta-function definition of the
determinants, consider the quantity
\beq
F(A,B)=\frac{\det (AB)}{(\det A)( \det B)}=e^{a(A,B)}
\:.\label{a4}\eeq
Introduce then the family of  $\Psi$DOs
\beq
A(x)=\eta^x B^{\fr{a}{b}}\,, \hs \eta=A^{b}B^{-a} \,,
\eeq
and define the function
\beq
F(A(x),B)=\frac{\det (A(x)B)}{(\det A(x))( \det B)}
\:.\label{a5}\eeq
One gets
\beq
F(A(0),B)=\frac{\det B^{\fr{a+b}{b}}}{(\det B^{\fr{a}{b}})( \det B)}=1
\,, \hs F(A(\fr{1}{b}),B)=\frac{\det (AB)}{(\det A)(\det B)}=F(A,B)
\:.\label{a6}\eeq
As a consequence, one  is led to deal with  the following expression
for the anomaly
\beq
a(A(x),B)=\ln F(A(x),B)=-\lim_{s \to 0} \ \partial_s \aq \Tr (
A(x)B)^{-s}-\Tr A(x)^{-s}-\Tr B^{-s} \cq
\:.\label{a7}\eeq
This quantity has the properties: $a(A(0),B)=0$ and $a(A(\fr{1}{b}),B)=
a(A,B)$.

The next step is to
compute the first derivative of $a(A(x),B)$ with respect to $x$,
the result being
\beq
\partial_x a(A(x),B)=\lim_{s \to 0} \ \partial_s \aq \Tr \at
\ln \eta \eta^{-xs} B^{-s\fr{a+b}{b}}\ct -  \Tr \at
\ln \eta \eta^{-xs} B^{-s\fr{a}{b}}\ct \cq
\:.\label{a8}\eeq
Making now use of Eq.~(\ref{a3}), one  obtains
\beq
\partial_x a(A(x),B)&=&C \frac{\mbox{res} (\ln \eta)}{b}-
 x \frac{\mbox{res} [(\ln
\eta)^2]}{a+ b}
-C \frac{\mbox{res} (\ln \eta)}{b}+ x \frac{\mbox{res} [(\ln
\eta)^2]}{a} \nn \\
&=&x  \frac{b}{a(a+b)} \mbox{res} [(\ln \eta)^2]
\:.\label{a9}\eeq
And, finally, performing the integration with respect to $x$,
 from $0$ to $1/b$, one
gets Wodzicki's formula for the multiplicative anomaly,
used in Sect. 3,  namely
\beq
a(A,B)=a(B,A)=\frac{\mbox{res}\aq (\ln(A^bB^{-a}))^2 \cq}{2ab(a+b)}
\:.\label{a10}\eeq


\end{document}